\documentclass[prd,aps,showpacs,preprintnumbers,amssymb]{revtex4}
\usepackage{axodraw}
\usepackage{color}
\usepackage{epsf}

\def\e3p{$\eta \rightarrow 3 \pi$}

\begin{document}
\title{%
\hfill{\normalsize\vbox{%
\hbox{}
 }}\\
{ Phase transitions for SU(N) gauge theories with arbitrary number of flavors}}

\author{Renata Jora
$^{\it \bf a}$~\footnote[1]{Email:
 rjora@theory.nipne.ro}}

\affiliation{$^{\bf \it a}$ National Institute of Physics and Nuclear Engineering PO Box MG-6, Bucharest-Magurele, Romania}

\date{\today}

\begin{abstract}
We study the phase diagram of an $SU(N)$ gauge theory in terms of the number of colors $N$ and flavors $N_f$ with emphasis on the confinement and chiral symmetry breaking phases. We argue that as opposed to SUSY QCD there is a small region in the $(N,N_f)$ plane where the theory has the chiral symmetry broken but it is unconfined. The possibility of a new phase with strong confinement and chiral symmetry breaking is suggested.
\end{abstract}
\maketitle

\section{Introduction}

Phase diagram of a nonabelian gauge theory with almost massless fermions has been the subject of many intensive studies \cite{Banks}-\cite{Schafer} over the past decades.
 The phase structure of supersymmetric QCD  in terms of the number of flavors and colors has been elucidated in \cite{Seiberg}. Although it is known that in QCD phases like confinement and chiral symmetry breaking are present in the phase diagram, their exact occurrence is far from being settled. Quite a few questions regarding the phase diagram of QCD are actual \cite{Schafer}, especially when one makes a comparison to the supersymmetric counterpart.

In this paper we shall adopt a new point of view with regard to the phase structure of QCD with fermions in the fundamental representation based on the knowledge that the potential between two sources is one of the indicators of the phase in which the theory is in. In order to establish the respective behavior we rely on the study of the Callan Symanzik equation for the two point gluon Green function. In this context we will be able to shed a new light on the possible phases and especially on the passage to the confinement and chiral symmetry breaking ones.

First let us briefly review what it is the current view with regard to the QCD phase diagram.
We start from an $SU(N)$ gauge theory with $N_f$ fermions in the fundamental representation. Beta function for this theory is known up to the fourth order \cite{Larin}. In the 't Hooft scheme \cite{Hooft1}, \cite{Hooft2}  beta function stops at the first two orders coefficients. This part of the beta function is also renormalization scheme independent so provides a useful framework to study the behavior of the coupling constant.  In \cite{Jora} by studying the global properties of the partition function it was proved that in a sense the two loop beta function,
\begin{eqnarray}
&&\beta(g^2)=\frac{d g^2}{d\ln\mu^2}=-b\frac{g^4}{16\pi^2}-c\frac{g^6}{256\pi^4}=
\nonumber\\
&&-[\frac{11}{3}N-\frac{2}{3}N_f]\frac{g^4}{16\pi^2}-[\frac{34}{3}N^2-2\frac{N^2-1}{2N}N_f-\frac{10}{3}N N_f]\frac{g^6}{256\pi^4}.
\label{beta435}
\end{eqnarray}
contains the main  properties of the gauge coupling constant. This is due to the fact that for any function of $g^2$ that expanded in $g^2$ respects the first two order renormalization scheme independent coefficients there is one renormalization scheme for which that function corresponds to a beta function for the gauge coupling constant. Here one makes the underlying assumption that a series expansion in $g^2$ makes sense and thus $g^2$ is small. For the nonperturbative regime or the limit between  the perturbative and nonperturbative regions there are techniques that one may address. In the present work we adopt an approach based on the extension of the validity of the Callan-Symanzik equations to the nonperturbative domain.  However some important indicators about the QCD phase diagram come from the beta function at two orders, no matter which renormalization scheme is chosen.

We observe that for $b>0$, $N_f<\frac{11}{2}N$ the theory is asymptotically free which means that the coupling constant goes to zero for very large momenta. For $b<0$ the theory loses its asymptotic freedom and it is  infrared free. Since for $b<0$ also $c<0$ this property is preserved at two loops.

Let us discuss in more detail the region where $b>0$.  As one decreases the number of flavors from $N_f=\frac{11}{2}N$ first the coefficient $c<0$.  This means that there is a solution to the equation $\beta(g^2)=0$ so the beta function has a non trivial fixed point. This occurs for ($\alpha=\frac{g^2}{4\pi}$):
\begin{eqnarray}
&&-b\alpha^{*2}-\frac{c}{4\pi}\alpha^*=0
\nonumber\\
&&\alpha^*=-\frac{4\pi b}{c}.
\label{crit567}
\end{eqnarray}
It can be shown \cite{Appelquist1}, \cite{Appelquist2} that for $N_f$ close to $\frac{11}{2}N$  the coupling constant is small and one can solve from the beta function for it to obtain that for small momenta the coupling constant approaches the fixed point. The corresponding phase is called the conformal phase and has been described in the literature \cite{Miransky1}. As $N_f$ decreases even more the coupling constant increases and at some value of $N_f<\frac{11}{2}N$ confinement and chiral symmetry breaking occur. Although it is known that these two phases coexist for lower values of $N_f$ it is still debatable if the phase transitions for both of them occur at the same critical value of $N_f$. We know that for SUSY QCD there are regions where the theory is confining but still chirally symmetric.  Chiral symmetry breaking happens according to several studied \cite{Maskawa}, \cite{Fukuda}, \cite{Fomin},\cite{Sannino} when,
\begin{eqnarray}
\alpha=\alpha_c=\frac{2\pi N}{3(N^2-1)}.
\label{crit6665}
\end{eqnarray}
or when the anomalous dimension of the fermion mass operator $\gamma_m=1$ (see the next sections for a discussion of this point).

One can further use $\alpha_c=\alpha^*$ to determine the critical number of flavors where chiral symmetry breaking takes place:
\begin{eqnarray}
N_f^c=N\frac{100 N^2-66}{25N^2-15}.
\label{crit123}
\end{eqnarray}

Thus for $N_f^c<N_f<\frac{11}{2}N$ the theory is in the conformal phase; has a fixed point, is chirally symmetric and deconfined. For $N_f<N_f^c$ the theory is confined and with the chiral symmetry broken.

\section{Confinement and chiral symmetry breaking}

It was shown that in SUSY QCD \cite{Seiberg} there is a window $N_c+1<N_f<3N_c$ where the theory is asymptotically free but not confining. Also for $N_f=N_c+1$ the theory displays confinement but not chiral symmetry breaking. In the following we will claim that as opposed to SUSY QCD in QCD chiral symmetry breaking takes place before confinement.

We shall start from the Callan Symanzik equation for the two point gluon Green function:
\begin{eqnarray}
[p\frac{\partial}{\partial p}+2-\beta(g)\frac{\partial}{\partial g}-2\gamma_3-\gamma_m m\frac{\partial}{\partial m}]G^2(p,m,g)=0.
\label{eq221}
\end{eqnarray}
Here $p$ is the momentum, $\beta(g)$ is the beta function, $\gamma_3$ is the anomalous dimension of the gluon wave function and $\gamma_m$ is the anomalous dimension of the fermion mass operator.
We need to clarify what are the exact definitions of the beta function and anomalous dimension that should be introduced in Eq. (\ref{eq221}). We will work in the background gauge field method where $Z_3$ is the renormalization constant for the gluon wave function. Then:
\begin{eqnarray}
&&\beta(g)=\frac{1}{2}g\frac{\partial Z_3}{\partial \ln\mu}
\nonumber\\
&&\gamma_3=-\frac{1}{2}\frac{\partial Z_3}{\partial \ln\mu}=-\frac{\beta(g)}{g}.
\label{def3443}
\end{eqnarray}

For the correct definiton of the other anomalous dimensions that enter the Callan Symanzik equation one takes into account the positive shift of the corresponding quantity as opposed to the standard dimensional regularization definition where one takes into account the negative one. Thus the correct expression for $\gamma_m$ in Eq. (\ref{eq221}) is,
\begin{eqnarray}
\gamma_m=-\frac{1}{m}\frac{\partial m}{\partial \ln\mu}.
\label{res3332}
\end{eqnarray}

We are interested in the behavior of the two point Green function at low energy. Without any loss of generality we can take $p=km$ where $p$ is the momentum, $m$ is the fermion mass (we consider  a generic mass for all fermions) and $k$ is an adimensional scaling constant. Then the Callan Symanzik equation (\ref{eq221}) becomes:
\begin{eqnarray}
[p \frac{\partial}{\partial p}(1-\gamma_m)+2-\beta(g)\frac{\partial}{\partial g}-2\gamma_3]G^2(p,g,m)=0.
\label{newcs34}
\end{eqnarray}
In the background gauge field method (see Eq. (\ref{def3443})) this can further be simplified to:
\begin{eqnarray}
[p \frac{\partial}{\partial p}(1-\gamma_m)+2-\beta(g)(\frac{\partial}{\partial g}-\frac{2}{g})]G^2(p,g,m)=0.
\label{cs445566}
\end{eqnarray}

We thus have a differential equation dependent on two variables $p$ and $g$ and we expect critical behavior when the different coefficients of the differential operators change the sign.
First we need a solution of the two point function of the Eq. (\ref{cs445566}) at the infrared fixed point. We thus ask $\beta(g)=0$ (for which $\frac{g^{*2}}{4\pi}=-\frac{4\pi b}{c}$) which leads to:
\begin{eqnarray}
[(1-\gamma_m)p\frac{\partial}{\partial p}+2]G^2(p,g,m)=0.
\label{rez32321}
\end{eqnarray}
This equation has a simple solution:
\begin{eqnarray}
G^2(p,g,m)\approx \frac{1}{p^{\frac{2}{1-\gamma_m}}}.
\label{sol787}
\end{eqnarray}
We require $\gamma_m=2$, where at one loop:
\begin{eqnarray}
\gamma_m=3\frac{N^2-1}{N}\frac{g^2}{16\pi^2},
\label{g554}
\end{eqnarray}
and solve for the corresponding value $g_c$:
\begin{eqnarray}
\frac{g_c^2}{4\pi}=\frac{8\pi N}{3(N^2-1)}.
\label{cr4566}
\end{eqnarray}
Around this particular value for $\gamma_m$ the Green function $G^2(p,g,m)$ will be close to:
\begin{eqnarray}
G^2(p,g,m)\approx p^2,
\label{rez434}
\end{eqnarray}
which for low momenta and in the coordinate space will lead to a confining potential of the type $V(r)\approx r$ (see \cite{Jora1} for details). We claim that this is a clear indicator that the transition to the confinement phase takes place around this point. Since we are already are in the critical regime where $g=g^*$ we further require $g_c=g^*$ to obtain the corresponding number of flavors:
\begin{eqnarray}
N_f^c=N\left(\frac{101N^2-33}{32N^2-12}\right).
\label{crit6657}
\end{eqnarray}

Note that this is in slight contradiction with previous studies (see section I) which claim that the transition to confinement take place at the value in Eq. (\ref{crit123}).

Early studies \cite{Georgi}  suggest that the behavior of the mass around the chiral symmetry breaking point is:
\begin{eqnarray}
m(\mu)\approx\frac{1}{\mu}
\label{g665}
\end{eqnarray}
which in our notation means $\gamma_m=-\frac{d\ln m}{ d\ln \mu}=1$.
For $\gamma_m$ approaching $1$ from above we observe from  Eq. (\ref{rez32321}) that the Green function becomes zero for low momenta and infinite for high momenta  indicating again a critical behavior. We claim in agreement with \cite{Georgi} that this behavior of the Green function indicates the transition to the chiral symmetry breaking phase. The condition $\gamma_m=1$ leads to the critical coupling constant $\alpha_s$ ($g_s$):
\begin{eqnarray}
\alpha_s=\frac{g_s^2}{4\pi}=\frac{4\pi N}{3(N^2-1)}.
\label{rez43553}
\end{eqnarray}
Furthermore the constraint $g_s=g^*$ yields the critical number of flavors for the transition to the chiral symmetry breaking phase:
\begin{eqnarray}
N_f^s=N\left(\frac{67N^2-33}{19N^2-9}\right).
\label{cr5454}
\end{eqnarray}
Note that this value is still in the region where $b>0$ and $c<0$ where the non trivial fixed point exists.

For $N$ large the transition to the chiral symmetry breaking phase happens at $N_f^s \approx 3.5 N$ whereas the confinement transition occurs at $N_f^c\approx 3.1N$ showing that for certain values of N there are values of $N_f$ for which, as opposed to the supersymmetric case,  there is chiral symmetry breaking but no confinement.  For example for $N=3$, $N_f^c=9.5$ and $N_f^s=10.5$ and for $N=4$, $N_f^c=12.7$ and $N_f^s=14.08$ thus proving our point. More specific  for $N=3$ and $N_f=10$ there is chiral symmetry breaking and no confinement as confinement sets in at $N_f=9$. Also for $N=4$, $N_f=14$ or $N_f=13$ there is chiral symmetry breaking  and no confinement as the latter commences at $N_f=12$.

To resume we showed that for an $SU(N)$ abelian gauge theory with fermions in the fundamental representation confinement sets in for slightly lower values of $N_f$ than chiral symmetry breaking and  thus there are region in the phase diagram in terms of $N$ and $N_f$  where there is chiral symmetry breaking but no confinement. This result can be related to opposite findings in supersymmetric QCD.

\section{Hints of other critical behavior}

According to the analysis in section II the two point Green function of the gluon suggest that in the region $b>0$ and $c<0$ there are only two main phase transitions associated to confinement and chiral symmetry breaking. For $b<0$  the system is in the free infrared phase. In this section we will try to find if there are signs of critical behavior in the region $b>0$, $c\geq0$. For that we rewrite the Callan Symanzik equation in the background gauge field method:
\begin{eqnarray}
[p\frac{\partial}{\partial p}(1-\gamma_m)+2-\beta(g)(\frac{\partial}{\partial g}-\frac{2}{g})]G^2(p,g,m)=0,
\label{newc554}
\end{eqnarray}
and we look for possible changes of sign in the differential operators in Eq. (\ref{newc554}) for $\beta(g)\neq 0$. We have two possibilities:
\begin{eqnarray}
&&(\frac{\partial}{\partial g}-\frac{2}{g})G^2(p,g,m)=0
\nonumber\\
&&(2+\frac{2}{g}\beta(g))G^2(p,g,m)=0.
\label{rez44343}
\end{eqnarray}

The constraint in the first line of Eq. (\ref{rez44343}) is in the perturbative regime and does not bring anything new. The condition in the second line leads to $\beta(g)=-g$ and to the simplified equation:
\begin{eqnarray}
[(1-\gamma_m)\frac{\partial}{\partial p}-\beta(g)\frac{\partial}{\partial g}]G^2(p,g,m)=0.
\label{rez4435}
\end{eqnarray}
This equation can be reduced to:
\begin{eqnarray}
p\frac{\partial G^2(p,g,m)}{\partial p}=\frac{g}{\gamma_m-1}\frac{\partial G^2(p,g,m)}{\partial g}.
\label{rez4434}
\end{eqnarray}
There is no much we can learn from the behavior of the Green function from the above equation (as it is dependent on the constants of integration) but we can spot a singularity for $\gamma_m=1$ which might indicate critical behavior. Since we were already in the regime with confinement and chiral symmetry breaking this possible critical behavior might indicate a transition to a phase or region where these two become strong.
The relations:
\begin{eqnarray}
&&\beta(g)=-g
\nonumber\\
&&\gamma_m=1,
\label{newcond5546}
\end{eqnarray}
lead to  the corresponding critical number of flavors:
\begin{eqnarray}
N_f^{cs}=\frac{40N^4+21N^2-27}{19N^3-9N}.
\label{rez2213}
\end{eqnarray}
However this tentative phase transition should be confirmed by alternative studies.

\section{Discussion and conclusion}

Phase diagram at zero temperature of a $SU(N)$ gauge theory with fermions in different representations has been discussed extensively in the literature \cite{Appelquist1}-\cite{Schafer}. In particular the critical number of flavors at which chiral symmetry breaking happens has been obtained using many approaches like the gap equation \cite{Appelquist1}, \cite{Appelquist2},  nonperturbative effective potential \cite{Sannino}, nonperturbative dynamics of the gauge fields \cite{Miransky}, lattice calculations \cite{Iwasaki}, Nambu Jona Lasinio model \cite{Gies}, Bethe Salpeter equation \cite{Harada}. In most of these cases the critical number of flavors obeys $N_f^c< 4N$ thus suggesting that many of the methods converge to a similar result.

If the regions of confinement and chiral symmetry breaking are confirmed than the rest of the phases that exist for $N_f>N_f^c$ are already settled and are determined entirely by the behavior of the beta function. Thus the infrared free phase corresponds to $N_f>\frac{11}{2}N$ where the theory loses its asymptotic freedom and the conformal window belongs to the region $N_f^c<N_f<\frac{11}{2}N$.

In this paper we introduced a new way to study some of the features of the phase diagram in the $(N,N_f)$ plane that relies only partially on the beta function. We start from what one can learn from the beta function and anomalous dimensions and then we use this knowledge to analyze the two point gluon Green function from the corresponding Callan Symanzik equation. We argue that these equations describe accurately the all order Green functions and thus contain hints about their nonperturbative behavior. Our approach is particularly useful to describe to transitions to the confinement and chiral symmetry breaking phase because we have some information about how the potential between two sources should behave for these cases. One of our main results is that that the transitions to chiral symmetry breaking should occur for,

\begin{eqnarray}
N_f^s=N\left(\frac{67N^2-33}{19N^2-9}\right),
\label{first554}
\end{eqnarray}
 whereas transition to the confinement phase happens for a slightly lower value for $N_f$:
\begin{eqnarray}
 N_f^c=N\left(\frac{101N^2-33}{32N^2-12}\right).
 \label{nerw34}
 \end{eqnarray}

Furthermore the study of the Callan Symanzik equation suggest that for values of $N_f$ even lower there is a possibility that a new phase transition occurs:
\begin{eqnarray}
N_f^{cs}=\frac{40N^4+21N^2-27}{19N^3-9N}.
\label{rez2213}
\end{eqnarray}

We suggest that this new phase transition  corresponds to a region where both confinement and symmetry breaking become strong as the two loop beta function $\frac{\beta(g)}{g}\leq-1$. However we could not extract more information with regard to the behavior of the system in this phase from the analysis we made.  A complete confirmation and description of this phase would require alternative nonperturbative techniques.

\begin{figure}
\begin{center}
\epsfxsize = 8cm
 \epsfbox{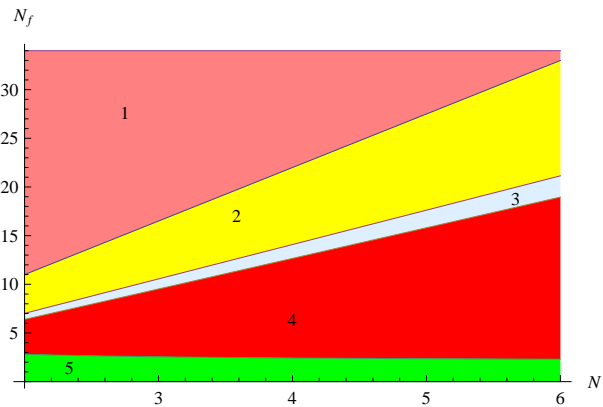}
\end{center}
\caption[]{%
Phase diagram of an $SU(N)$ gauge theory in terms of the number of colors $N$ and flavors $N_f$. The numbers from 1-5 represent the distinct phases:1-infrared free; 2-conformal; 3-weak chiral symmetry breaking; 4-confinement and chiral symmetry breaking; 5-strong confinement and chiral symmetry breaking. Note that chiral symmetry breaking phase corresponds to region $3$ but all subsequent confinement phases are also characterized by chiral symmetry breaking.
}
\label{phases}
\end{figure}

 In Fig \ref{phases} we compile all the information in the present work and depict the full zero temperature phase diagram of a $SU(N)$ gauge theory with fermions in the fundamental representation.

\section*{Acknowledgments} \vskip -.5cm

The work of R. J. was supported by a grant of the Ministry of National Education, CNCS-UEFISCDI, project number PN-II-ID-PCE-2012-4-0078.

\end{document}